# Jamming of molecular motors as a tool for transport cargos along microtubules.

Lucas W. Rossi and Carla Goldman

Departamento de Física Geral - Instituto de Física

Universidade de São Paulo CP 66318

05315-970 São Paulo, Brazil.

January 2012

## Abstract

The hopping model for cargo transport by molecular motors introduced in Refs. [12, 13] is extended in order to incorporate the movement of cargo-motor complexes (C-MC). Hopping processes in this context expresses the possibility for cargo to be exchanged between neighboring motors at a microtubule where the transport takes place. Jamming of motors is essential for cargos to execute long-range movement in this way. Results from computer simulations of the extended model indicate that cargo may indeed execute bidirectional movement in the presence of motors of a single polarity, confirming previous analytical results. Moreover, these results suggest the existence of a balance between cargo hopping and the movement of the complex that may control the efficiency of cargo transfer and cargo delivery. An analysis of the energy involved in this transport process shows that the model presented here offers a considerable advantage over other models in the literature for which cargo dynamics is subordinated to the movement of the C-MC.

# 1 Introduction

Cargo particles move bidirectionally as they are transported by molecular motors along microtubules. A current explanation of this phenomenon, expressed by the so called coordination model, relies on the idea that motors of different polarities are coordinated by external agents to work on the same particle at different times. In a related explanation, the tug-of-war model, the two kinds of motors would act simultaneously by pulling the cargo to one or to the other side of the microtubule [1, 2, 3, 4].

A general difficulty encountered in any of these views concerns the presence of other particles on the microtubules, including other motors, that may impose restrictions on cargo´s motion. In fact, as noticed in Ref. [5], there are diverse "physical barriers" at the cytoskeleton where intracellular transport takes place. The cytoskeleton itself consists of a highly structured composition of crossed filaments on which there are present associated proteins and other motors (and other cargos) that may limit both motor and cargo´s motion [6]. Because of this, the origins of the bidirectional movement of cargo, including organelles, vesicles, virus and other particles, on microtubules is still a matter of intense debate [7]. Other models have been proposed in the literature as improvements on the coordination or tug-of war models and are formulated by attributing a dynamic role to the microtubules due to their elastic properties and intrinsic polarity [8, 9, 10, 11]. Nonetheless, a more complete consideration of questions related to the traffic of motors in these contexts is still required.

The occurrence of motor jamming on crowded microtubules would impose difficulties to the coordination-like models even if there were present motors of just a single polarity. In fact, descriptions of the transport phenomenon in such contexts are based on the premise that cargos can move only if attached to motors arranged to form a cargo-motor complex (C-MC). Each C-MC is supposed to follow the dynamics of the constituting motors. We shall refer to these as C-MC models. As noted in Ref.[11], considering even that motors may eventually be detached from and then reattached to microtubules in order to temporarily create more space for the C-MC´s, it is not clear how this would help the system to achieve the expected efficiency in the transport process. Motor attachment and detachment occur at random positions on the microtubules, not necessarily at the places or times that would be required for cargo passage.

Another related problem concerns the nature (type, strength, etc.) of the bond (linkage) between cargos and motors. In carrying cargos along relatively long distances, it would be necessary for C-MC complexes to maintain a stable attachment between cargo and motor as they move along tracks. On the other hand, it would seem that a strong attachment between the two particles in this context would restrain cargo release from motors at the required places and at the right times.

Thus, the reality of traffic jamming and the mechanisms through which local coordination might be achieved still challenge the current views of the transport processes based on C-MC models.

Motivated by this, we have been exploring the problem from a different perspective as an alternative to the idea that C-MC assembly is a necessary condition for transporting cargo in this context [12, 13]. According to this alternative view, cargo transport would result from a sequence of elementary hopping processes taking place on a microtubule represented by a one-dimensional lattice. Introduced in the pioneering work by Kolomeisky and Widom [14], one-dimensional hopping models have ever since been used to describe the dynamics of molecular motors along microtubules. Many adaptations of the original idea have contributed to unravel details of the phenomenon and specially, the collective character of the related processes. Although representing simplified descriptions of the reality, it is believed that these 1D hopping models capture essential and relevant features of motor dynamics. It should not be expected, however, to obtain from them detailed quantitative predictions about the system. Using a reasonable number of parameters, stochastic models of this sort are expected to offer restricted although important proposals regarding the physical mechanisms of interest. In a sense, our model extends the original idea to account also for cargo hopping in addition to the underlying motor hopping. It was conceived originally on the following basis:

(i) Motors and cargos would not assemble to form stable C-MC complexes. A weak and intrinsically flexible attachment (or "floppy linkage", as coined in Ref. [15]) that might eventually be established between motors and cargos would be short-lived. The relevant degrees of freedom of such transiently assembled structures would be excited by thermal fluctuations (noise).

(ii) Because of these thermal fluctuations, cargo may be exchanged (or "hop") between motors occupying neighboring sites on the lattice.

It is worth mentioning here that both elements, namely thermal fluctuations and cargo exchange have already been observed in experiments. Fluctuations in the relative positions of cargo and associated motors have been detected as they introduce difficulties in characterizing experimentally the movement of motors by following the movement of the cargo [16]. Cargo exchange (or cargo switch, or cargo hop, as we call here) between motors of different polarities, moving on different structures like actin filaments or microtubules, has been observed *in vivo* [5]. Actually, cargo switching was found to be a useful mechanism to move cargo across the diverse structures of cytoskeleton . Therefore, the scheme in Fig.1 expresses the idea of combining (i) and (ii) in order to examine their effect on cargo transport, restricted to a one-dimensional space, taking place on crowded microtubules and involving motors of a single polarity.

We have shown in our previous studies that long-range movement of cargo may be achieved in this way if (and only if) motors become jammed. Cargo would then be able to move through long distances as it undergoes a sequence of these elementary (short-range) steps, hopping from motor to motor, either forwards or backwards. Thus, in this view, and contrarily to common expectations, motor jamming along microtubules would not impede cargo flow. On the contrary, jamming would be desirable, as a condition for the whole process to attain a relatively high degree of efficiency.

Originally, the stochastic lattice model proposed in [12] and extended in [13] to explain the observed bidirectional cargo movement was conceived on the basis of ASEP models (asymmetric simple exclusion processes) already formulated in these contexts to describe the dynamics of a collection of single polarity motors [17, 18]. Because our interests focus on the study of mechanisms responsible for transport carried out by motors, it was necessary to include cargos and their interactions with the motors on the same track. To that end, we have made a few assumptions in order to define the nature of such multiparticle interacting system in conformity with (i) and (ii) above:

(I) existence of steric interactions among particles;

(II) restrictions to cargo movement if not by hopping process;

(III) restrictions to motor movement if attached to cargos;

Both (II) and (III) ensure that the C-MC´s are immobile in this model. Indeed, one expects that the presence of cargos on the microtubules affects motor motility. In turn, changes in motor motility should affect the transport of cargos. The hopping model accounts explicitly for this interplay and offers a way to examine the conditions for long-range cargo transfer in different contexts. Observe that the analysis of the properties of such a model must necessarily be of a global nature since the relevant phenomenon investigated is motor jamming which is intrinsically non-local. From such analysis, we have concluded that the bidirectional movement of cargo can indeed be achieved through hopping under jamming conditions even in the presence of motors of a single polarity, but only if more than one cargo participate in the dynamics [13]. As jamming takes place, a given cargo may become able to hop over large clusters assembled either behind it or at the back end of a cargo in front, covering in this way relative long distances in both directions. We then suggested that the conditions for these events may be controlled by adjusting the density of motors (number per unit volume) attached to the microtubule. Accordingly, no external agents would be necessary to determine the direction of cargo movement.

Here, we extend the hopping model by adding an extra process to the original dynamics. We confer to motors the ability to move to a neighbor lattice site even if attached to cargo. This means that we incorporate into the model the idea of the movement of the C-MC complexes. From a formal point of view, this recovers the ergodicity of the model, a question raised in Ref.[13]. In practical applications, this would allow one to investigate the effects on transport properties of this combination of two processes - the one dependent on C-MC complexes and the other based only on hopping. Yet we maintain the choice regarding the presence of single polarity motors in the system. Computer simulations of this extended model show that bidirectionality of cargo may result from this combination of processes even if there were present only a single cargo on the track. Estimates of related energy costs indicate that hopping may introduce significant advantage over mechanisms that rely exclusively on the movement of C-MC complexes.

The paper is outlined as follows. The original hopping model is briefly reviewed in Section 2. In Section 3 we present results for cargo displacement and average velocity obtained by computer simulation of the extended version. Energy estimates and final remarks are in Section 4.

## 2 The hopping model of cargo transfer

The original stochastic lattice model represented in **Figure 1(a-c)** has been mapped into an ASEP (asymmetric simple exclusion process) [19], [20], [21] for describing the following elementary processes that take place on a one-dimensional lattice (microtubule):

$$
\begin{aligned}
&\textbf{(a)} \quad 10 \rightarrow 01 \quad \text{with rate } k, \quad \text{probability } kdt \\
&\textbf{(b)} \quad 12 \rightarrow 21 \quad \text{with rate } w, \quad \text{probability } wdt \\
&\textbf{(c)} \quad 21 \rightarrow 12 \quad \text{with rate } p, \quad \text{probability } pdt
\end{aligned}
\tag{1}
$$

Label 1 is assigned to a site of this lattice that is occupied by a motor carrying no cargo; label 2 is assigned to a site occupied by a motor weakly attached to a cargo; and a label 0 is assigned to an empty site. Notice that the above dynamics preserves the number of motors and cargos on the lattice. In principle, fast processes describing motors attachment and detachment from the microtubule could be added to this as for example in $10 \leftrightarrows 00$ with appropriate rates. Such processes, however, should not modify the general characteristics of results presented below reached in the limit of very large number (or average number) of motors on the microtubule at stationary conditions. Because of this, we decided to keep the model as simple as possible in order to capture its essential features and understand the effects of C-MC movement on the already considered hopping process.

Process (a) represents an elementary step of a biased motor as it moves forward, towards the microtubule minus end according to the convention adopted here (Fig.1(a)). To define motor stepping is, of course, essential in building the model dynamics taking place on the microtubule since it is the primary source of jamming. This in turn may create the conditions necessary to the long-range transport of cargo at high motor densities. It is exactly this possibility that we wish to investigate here. Processes (b) and (c) represent the exchange of cargo between neighboring motors, to the left and to the right, respectively **(Fig. 1(b,c))**. Notice that each of these elementary steps occurs with a certain probability and under certain conditions. For process (a) to occur with the indicated probability it is required that the site to the right of the motor stays empty during the time interval $dt$. The other two processes depend on the presence of a motor to the left (b) or the right (c) of the motor attached to the cargo within $dt$. The stationary properties of this model are derived in [12] and [13], in the limit for which the number $n_1$ of motors and the number $N$ of sites on the microtubule are both very large in such a way that the ratio $n_1/N \to \rho$, i.e. converges to a finite density $\rho$ of motors. The analysis performed there focuses on the behavior of cargo average velocity $v_m$. For a broad range of values for the parameters, $v_m$ presents two distinct behaviors as $\rho$ varies, characterizing the occurrence of a phase transition in this system. Moreover, in cases for which there are present more than one cargo on the lattice, as considered in [13], $v_m$ changes sign after relative long runs. Thus, in contrast with a local *coordination* or local dispute conceived in the context of *tug-of-war* models, the phenomenon predicted in [13] emerges from the global properties of the system, related to the traffic and associated clustering of motors, which, in turn, can be controlled by tuning the amount of motors bound to the microtubule.

# 3 Combining hopping with the movement of the complex.

The idea here is to relax the condition used both in [12] and [13] under which the movement of cargo would take place exclusively through hopping. Accordingly, we shall add to the above dynamics the following process

$$(\mathbf{d}) \quad 20 \rightarrow 02 \quad \text{with rate } \alpha, \quad \text{probability } \alpha dt \qquad (2)$$

in order to let cargo to move also by means of a C-MC complex. Consistent with the fact that we have assumed the presence of motors of a single polarity, the complex shall be biased so as to move in a single direction, the same as that of the motors in (a). In general, however, the numerical values of $\alpha$ and $k$ need not to be the same. In fact, in a recent study using Monte Carlo simulation, it was concluded that an attached cargo can indeed modify significantly the rates at which motors bind to the microtubule, especially at high viscosities [22].

We also observe that as in our original model, the attachment between cargo and motors should be weak in order to allow cargo to be exchanged between neighboring motors. Notice that the idea of combining the movement of the complex with cargo hopping does not diminish the relevance of the traffic jam in this context. As we shall argue below, hopping and jamming conditions continue to play a crucial role in explaining the long-range movement of cargo, especially at high motor densities.

## 3.1 Numerical Study

We consider the extended hopping dynamics taking place at a one-dimensional lattice of $N$ sites along which $n_1$ motors and $n_2$ cargos, with $n_2 \leqslant n_1 < N$, are initially distributed at random. The time evolution of the system is then carried out with the aid of computer simulation through a sequence of *global runs* considering periodic boundary conditions. Within a global run, the $N$ sites, one at time, are tested for updating. The procedure is made sequential and the sites are selected at random. If a selected site, say $j$, has not already been updated during the run, then an attempt shall be made to interchange its occupancy with site $j+1$ according to the rules set in (1) and (2). If, however $j+1$ has already been updated during this run, then $j$ would remain unchanged. Subsequently, a new site is selected at random and the process is repeated until all sites are tested, which ends the run. A new run starts with its initial condition set by the final configuration attained in the previous run. The time unit $\Delta t$ is defined as one global run. The total number $T^\infty$ of global runs sets the time interval for evaluating the average values for the quantities of interest at stationary conditions. $T^\infty$ is a parameter of the algorithm. We seek stationary conditions by repeating the entire procedure with an increasing number of runs until the average profiles become invariant.

When the simulation starts, one of the cargos in the system - the one whose properties will be evaluated - is selected at random. At fixed values of the parameters, the movement of this selected cargo is marked at the end of each global run as $-1, +1$ or $0$ to indicate that it executed, respectively, a step to the right, to the left or not changed its position with respect to the previous run. The algebraic sum of all of these steps performed along the set of runs accounts for the total displacement $d(T^\infty)$ of the selected particle within each defined time interval $T^\infty$. Cargo average velocity $v_m$ is then estimated as the ratio $d(T^\infty)/T^\infty$.

**Fig.2** shows the results obtained in this way for the variation of $v_m$ as a function of $\rho$, at fixed values of parameters $k, \alpha, w, p$, and $n_2$, as indicated. The choice of parameters in each of these examples was not guided by pre-existing experimental data. Our main interest here is simply to understand the behavior of the model, specially regarding the relative contribution of each of the two modes considered to promote cargo movement. This allows us to identify the origins of some of the observed properties as, for example, the fact that $v_m$ may change sign as $\rho$ varies. This confirms our predictions made elsewhere, based on analytical calculations of $v_m$ using the model in [13] [1]. This particular result suggests that motor density at the microtubule may indeed play an important role as a control parameter to set cargo´s direction and thus the ability to change the course of its movement along the considered microtubule. **Fig.3** offers a more complete view of the behavior of $v_m$ with respect to a broader region of model parameters, at fixed number of cargos. Parameters $p$ and $w$ are both related to oscillations of the attached cargo with respect to the motor´s main symmetry axis. Therefore, if a bound cargo is able to induce a change onto motor with respect to its symmetry axis, it is conceivable that such change might well be represented through a choice of numerical values for these parameters such that $p \neq w$ (**Fig.3a**). On the other hand, the reasoning behind a choice that sets $k \neq \alpha$ has already been mentioned above. It is based on studies of the effects of viscosity on the motor motility in the presence of an attached cargo [22]. We must emphasize, however, that in spite of these possibilities, we notice in **Fig. (3b)** that it is not necessary to have $p \neq w$ neither $k \neq \alpha$ in order to observe changes in the sign of $v_m$ at varying values of $\rho$. Although it becomes clear in these figures that in case $k = \alpha$ (**Fig.3b**) the region of densities within which the signal of $v_m$ remains unchanged becomes larger than the corresponding region in case $k \neq \alpha$, still there are uncountable possibilities for cargo to reverse its direction of movement, either by changing $\rho$ or by changing $p$ (or $w$). In other words, a choice of parameters such that $p \neq w$ and/or $k \neq \alpha$ is not a necessary condition for our model to describe the bidirectional movement because it depends mainly on clustering, a phenomenon displayed by ASEP models even in the presence of a single type of particles (for example, in the absence of cargos).

---

[1] We have introduced in [13] a procedure to compensate for the lack of ergodicity, as the movement of the C-MC complexes is not considered explicitly. Such a procedure, however, does not introduce drifts to cargo movement. Thus, the characteristics of the long-range displacements predicted there are due exclusively to motor clustering.

Data in **Fig.2** can be better appreciated with the aid of the accompanying cargo displacement profiles $d(t)$ for $t \sim T^\infty$ measured in units of global runs $\Delta t$. These are shown in **Fig.4** for the same set of model parameters used to evaluate $v_m$ in **Fig.2(a)**, as indicated, for different choices of motor densities. The observed long-range displacements in each direction result from an interplay between two processes. One of those is motor clustering that enables cargo to execute long-range movements by hopping to both directions, and it is predominant at high motor densities. The other process is related to pure C-MC movements. It allows cargo to move steadily in the forward direction if there were no impediments on the microtubule; thus, it is predominant at sufficient low motor densities. Nevertheless, the results achieved here suggest that both processes play important roles at all motor densities. In fact, at high motor densities C-MC dynamics provides cargo with a mechanism to overcome the empty spaces between clusters and reach the next cluster so that it can resume its hopping-based movement. On the other hand, at low densities hopping allows cargo to overcome the problem of having a low number of motors or clusters of motors already assembled, in order to resume its C-MC based movement.

The examples of **Fig.4** illustrate these possibilities. **Fig.4(i)** displays the trajectory of the cargo under consideration at relatively low motor densities. Within this region it develops a straight movement, i.e. toward the forward direction (microtubule minus end) at a near constant average velocity. As just mentioned, this is likely to be due mainly to the C-MC-based movement. In fact, as shown in **Fig 5(i)**, the corresponding average size of the assembled clusters at such low motor densities is very small compared to the typical sizes of clusters assembled at higher densities **Fig 5(ii-iv)**. Therefore, hopping is not expected to contribute to the observed long-range movement within this region. As the density of motors increases, cargo decreases its velocity. Clustering then begins to contribute as a mode of cargo transport leading it to display forward as well as backward movements as it is able to hop over the small clusters **Fig 5(ii)** in both directions, as explained. Thus, at the point at which the average velocity $v_m$ becomes effectively zero, cargo movement is characterized by large fluctuations **(Fig 4(ii))** because then hopping would compete with the C-MC-based transport. This situation lasts until motor density becomes sufficiently large such that large clusters take over **(Figs. 5(iii))** enabling cargo to overcome long distances, this time through hopping. This explains the movement of the cargo toward the plus direction as the clusters are assembled behind it **(Fig 4(iii))**. At very high densities, once again cargo switches the direction of the drift **(Fig 4(iv))** which, in the considered situation, is likely to be due to hopping over large clusters that are assembled at the back end of another cargo present in the system. In this case C-MC dynamics just allows cargo to overcome the gap and reach the clusters in front. A similar analysis can be performed for the case shown in **Fig.2(b)** with a large number of cargos. The typical sizes of the assembled clusters in this case are much smaller than those shown in **Fig.5** (data not shown). Therefore, although hopping mode still operates at sufficiently high motor densities, specially along clusters at the back of a neighboring cargo, the movement should be imposed by that of the C-MC.

In view of this, we may suggest that changes of cargo´s drift direction in long-range displacements can be regulated by small variations in the density of motors attached to the microtubule under stationary conditions. This might explain the observed bidirectional movement in real systems.

## 4 Discussion

The hopping model for long-range cargo transfer by molecular motors is reviewed and extended in order to incorporate the dynamics of C-MC complexes. The results for the average cargo velocity obtained by numerical simulation indicate that the bidirectional movement displayed by cargo can be explained by this extended version of the model, even if there were in the system just a single cargo driven by single polarity motors.

Actually, this can be the case in real systems. Very recently, Roostalu et al. observed bidirectional motion of cargos in experiments performed in vitro with single type minus-end directed kinesin-5 Cin8 motor proteins [23]. Although the mechanisms that would trigger the phenomenon are not detailed in their paper, the suggestion made there is that it might be due to a reversal of Cin8 intrinsic polarity in situations in which many motors work together as a team.

We claim here that these new experimental findings can be accounted for by the hopping model with no changes to the properties of motors required. More precisely, if transport by single polarity motors takes place in the presence of noise that promotes cargo exchange, as explained above, then it would be possible to observe long-range movement of cargos in both directions. We have already predicted bidirectional movement through this mechanism in model systems possessing two or more cargos [13] . Here, we obtain similar results considering, in addition to hopping, unidirectional movement of just a single cargo through a C-MC interacting with the set of other motors present.

As noticed above, these two elements, namely cargo switching and noise have already been reported in the literature. Here we suggest a way to use them in order to built a model that describes the dynamics displayed by many interacting particles occupying the sites of a 1D lattice.

We then disclose the conditions under which motors assemble into relatively large clusters. These clusters, in turn, allow cargos to endure a sequence of such elementary hopping steps resulting in large displacements in either direction. It is known that ASEP models with only one type of particle undergo a dynamic phase transition at which clustering appears to be controllable by the particle density in the lattice [21]. We have shown that this also happens when cargos are added to the system. This allows us to conclude that 1) long-range cargo transport can be explained by the mechanism of hopping along such clusters, and also that 2) the relative amount (but not necessarily the polarity) of motors bound to the microtubule, i.e. the defined motor density parameter $\rho$, can control the direction of such large displacements. These conclusions come from the study of the behavior of average cargo velocity with respect to $\rho$ as depicted in **Fig.2.** In addition, the results suggest the existence of limiting values for motor densities to control transport operation. Cargo direction and therefore the effectiveness of cargo delivering would be self-regulated by small changes of motor density, especially if the system is close to the jamming transition.

Regarding this point, it is not clear to us how and even if the study performed in Ref.[23] at varying motor density in gliding assays can be compared to the results achieved here. Those studies focus on the properties of cargo-motor interactions; thus, in principle, the results could be used to investigate the magnitude of cargo fluctuations around a motor´s position as described here. On the other hand, the fact that the relatively large cargos considered in these experiments are not allowed to move through the C-MC introduces difficulties for a direct comparison between the data obtained there and the theory discussed here. Notice that due to their finite extension, the cargos considered in the experiments can indeed traverse the gaps between clusters of motors with no need for the C-MC mechanism. Yet, it is noticeable the similarity between the qualitative behavior shown in the experimental results and the predictions made here, within the considered motor density range. In any case, as argued by Roostalu et al., the quantity of motors at the microtubule seems to be an important tool for controlling cargo movement and direction. This is completely consistent with our previous claims [12, 13, 24] and it is emphasized by the results presented here.

In the data referred to above, one observes motor accumulation near the cargo being observed as it moves towards the plus-end side of the microtubule, in the opposite direction of individual Cyn-8 Kinesin motors. To explain these data the authors have suggested that i) the effect reflects some collective properties since motors work as a team to move the cargo, and ii) such collective property would then induce motors to change their intrinsic polarity. They concluded that the Cin-8 motors themselves may behave as bidirectional motors - individually, they would follow their minus-end intrinsic polarity, whereas, if working as a team, they would move and transport cargo according to the C-MC mechanism toward the plus-end direction. There is no attempt in their work to elucidate the mechanisms responsible for the alleged change in polarity.

We argue that there is another way to think about this data based on the ideas discussed here. In the context of the hopping model one does not require changes in individual motor polarity to explain the observed movement of the team of motors, although the relevant effect would indeed be attributed to collective properties developed by the system due to the global nature of the jamming process. Jamming depends on the dynamics and interactions of all motors and cargos present in the system, not just on the properties of the motors participating in the local team. Accordingly, we do understand why and how motors can accumulate next to cargo, as observed, just because the presence of cargo, although not being a necessary condition, enhances the conditions for motor jamming in its neighborhood.

The fact that a motor cluster can indeed move toward the opposite direction from that of the constituting motors may be better appreciated with the aid of **Fig.6**. It illustrates a situation in which a cluster is being formed. Motors moving toward the minus end would encounter the cargo and get jammed behind it. In turn, the presence of this cluster would induce cargo to step over it, toward the plus direction. Therefore, motors that were previously accumulated behind a cargo would pass to a position in front of it (because cargo moves back) and this would tend to disperse the cluster, as these motors, now free to move, would continue their movement toward the minus end of the microtubule. On the other hand, motors continuously reaching the cluster at its back end would tend to increase the cluster. Thus, at the same time that the cluster loses motors in front of it, it gains motors behind so as to appear that it is moving toward the plus-end direction, opposite to the intrinsic motor polarity. A balance between the tendency of losing motors and acquiring motors would eventually equilibrate a cluster´s size. In conclusion, the cluster (not the motors!) would appear to move to the plus-end direction due to a dynamic process of losing and gaining motors, but not because of changes in individual motor polarity. The cargo, on the other hand, if able to hop over the cluster, it may move in either direction, but the drift would be toward the plus-end, accompanying overall cluster movement.

Of course the dynamics exemplified above can be understood as a microscopic description of a *shock wave* [19] in this context, similar to the continuum version studied in Ref. [24]. We may then say that the hopping dynamics discussed here indeed expresses the relevance of the collective behavior of motors and cargos to the transport process and offers a novel description to the phenomenon. The results are simple, although nontrivial, in many respects, and they include a description of bidirectional effects.

Nonetheless in more realistic cases within the cell environment, one should not exclude the possibility of the presence of motors of both polarities on the same microtubule. Notice, however, that in the context discussed here, we understand that the presence of both kind of motors would simply enhance the conditions for motor jamming [25]. Thus, in principle, the presence of different motors would in fact create more possibilities for cargo to move in both directions but not necessarily as an effect of local coordination or competition between motors of different polarities, but instead, as a consequence of the combined effects that these two kinds of motors would have on traffic jamming and thus on motor clustering. Once again, this emphasizes the idea that the kind of long-range movement discussed here expresses collective effects involving all particles on the microtubule since jamming is not a local phenomenon.

Finally, we should notice that the mechanism discussed here does not require special stability of cargo-motor binding. On the contrary, the exchange of cargos would be facilitated both by an unsteady attachment between cargos and motors and, of course, by the flexibility of the motor tail.

## 4.1 Energy

It is interesting to estimate the energy cost $E_h$ associated with cargo transport in the context of the extended hopping model to compare with an equivalent quantity $E_{c-mc}$ for pure C-MC models. This can be done, for example, by estimating the energy required in each case to drive a cargo between two lattice sites that are far apart. For simplicity, we restrict the analysis to the case in which the movement of cargos takes place only in a definite direction, to the minus-end, say. This condition is accomplished in the context of the extended hopping model by setting $\omega = 0$ in (1). The corresponding condition in the context of C-MC models exists in the case in which single polarity motors are present. Our aim is to obtain lower and upper bounds to the quantity $E_h/E_{c-mc}$, the ratio between the corresponding energies in the two models.

Let $\delta_1$ and $\delta_3$ be the energies required in the processes $10 \to 01$ and $20 \to 02$, respectively, and let $\delta_2$ be the energy cost for exchanging a cargo between neighboring motors, process $21 \to 12$. Considering that the energy necessary for a motor protein to move one step forward is of the order of the energy released by the hydrolysis of one ATP molecule [26], we estimate $\delta_1 \sim 500 \cdot 10^{-21} J$. For $\delta_2$ we use the energy associated with thermal fluctuations needed for hopping. Accordingly, $\delta_2 \sim k_B T \sim 4,3 \cdot 10^{-21} J$ at body temperatures. It is more difficult to estimate $\delta_3$. Yet, because both $\delta_1$ and $\delta_3$ are related to the step of a motor we may consider that $\delta_3 \gtrsim \delta_1$.

The energy required in pure C-MC models is that for carrying a cargo along a distance comprising the whole set of $N$ sites, starting and ending at site 1 since the system presents periodic boundary conditions. Let $n_2$ be the total number of cargos to be transported and $n_1$ the total number of motors distributed along the $N$ sites. For simplicity, we suppose that cargos are allowed to attach to a single motor each, and also that there is present only one cargo in the system. Consequently, $n_2 = 1$ and $n = n_1 - 1$ is the number of motors bound to the microtubule at each time that carries no cargo. We now consider the possible configurations that may be reached by the system in a delivering process, starting from a configuration in which the $n$ motors are distributed in sequence between sites labeled $N-n+1$ and $N$. The C-MC starts at position $N+1 = 1$ following periodic boundary conditions. In this situation the energy cost to move the cargo by means of the C-MC exclusively would assume a minimum value. This is because the motors arranged in this way would need to move forward just along a minimum number $n+1$ of sites each, in order to provide enough space to the C-MC for reaching site 1 again, after completing the cycle. Any other initial arrangement would require that motors move along a larger number of sites than $n+1$. Then, the minimum energy $E_{c-mc}^{(0)}$ required for the complex to complete its way across the $N$ sites can be estimated as

$$E_{c-mc}^{(0)} \simeq \delta_1 n(n+1) + \delta_3 N \tag{3}$$

where the index (0) in $E_{c-mc}^{(0)}$ refers to the initial configuration under consideration. The first term at the RHS accounts for the energy to move the $n$ unbounded motors across $(n+1)$ lattice sites;

the term in $\delta_3$ accounts for the energy to move the complex with its cargo.

Starting from the same initial configuration, we are now able to determine upper and lower bounds for the corresponding energy $E_h^{(0)}$. In the context of the extended hopping model, cargo is also allowed to move by hopping, and thus spending less energy per step ($\sim \delta_2$) if compared to the movement through the C-MC ($\sim \delta_3$). A minimum amount $E_{h(\min)}^{(0)} = \delta_2 n + \delta_3 (N - n)$ of energy is required to complete the circle in cases for which cargo hops (instead of moving with the aid of C-MC) along the maximum number $n$ of unbounded motors. This is accounted for by the term in $\delta_2$. The term in $\delta_3$ accounts for the energy to move cargo along the unoccupied sites as it attaches to a motor to form a C-MC complex. If, however, all of the $n$ unbounded motors move in order to provide space to the C-MC, the energy involved in completing the circle would be exactly the same as $E_{c-mc}^{(0)}$ given by (3). This means that $\delta_2 n + \delta_3 (N - n) \leqslant E_h^{(0)} \leqslant E_{c-mc}^{(0)}$ or, in terms of $\delta_2/\delta_3 \equiv \varepsilon << 1$,

$$f_N(\varepsilon, \rho) \leq \frac{E_h^{(0)}}{E_{c-mc}^{(0)}} \leq 1 \tag{4}$$

where we have defined

$$f_N(\varepsilon, \rho) \equiv \frac{1 + (\varepsilon - 1)\rho}{1 + N\rho^2} \tag{5}$$

A remark is in order here in respect to the multiplicity of cases for which hopping may be combined with the C-MC movement. Within the pure C-MC context, any attempt by the complex to complete the path would necessarily involve energies equal to $E_{c-mc}^{(0)}$. When hopping is added to the dynamics, it creates a large number of possibilities for trajectories that can be followed by cargo, most of them accomplished by spending energies that are significantly less than $E_{c-mc}^{(0)}$. High energies would be required only in the very rare occasions in which the path is accomplished with no hopping or just a few events of hopping. **Fig.7** shows the behavior of the *gap*

$$g_N(\varepsilon, \rho) = 1 - f_N(\varepsilon, \rho) \tag{6}$$

between the upper and lower bounds expressed by Eq. (4) as $\rho$ varies, and at different values of $N$. At the scales being considered, the function $g_N(\varepsilon, \rho)$ is practically insensitive to variations of $\varepsilon$ in the range $0.01 \leq \varepsilon \leq 0.9$ (results not shown). Notice, however, that $g_N(\varepsilon, \rho)$ increases with $\rho$. This means that by increasing motor density, the number of possibilities for cargo to follow a path that requires less energy than $E_{c-mc}^{(0)}$ increases. There is an accompanying increase in the "entropy", namely in the number of different paths involving the same number of hops and the same number of C-MC steps that may be followed in different combinations. Thus, if the multiplicity of paths is accounted for, one might conclude that in the context of the extended hopping model, the events that do not involve hopping do not occur in practice. The curves in **Fig.7** suggest that, regarding energy costs, the two schemes - extended hopping and pure C-MC- become comparable only at artificially low motor densities .

# Acknowledgements

This work is supported by Fundação de Amparo à Pesquisa do Estado de São Paulo (FAPESP).

# Figure Caption

**Figure 1** - Dynamics of motors and cargos. (a) Step of a motor. The time spent by the motor with the two heads attached to the microtubule is much larger than the time it spends with just one of the heads attached to it [27]. This is part of the "hand-over-hand" mechanism proposed to explain the kinetics of two-headed motor proteins [26]. In view of this, we shall consider that occupation of a site by a motor occurs whenever it is occupied by the two heads. Cargo hopping occurs through a mechanism of exchange between neighboring motors. Due to the flexibility of the motor tails, the attached cargo may be caught either (b) by the motor at its right or (c) by the motor at its left. (d) Elementary dynamics of a C-MC complex.

**Figure 2** - Average cargo velocity $v_m$ as a function of motor density $\rho$. The parameters used are $N = 100$, $T^\infty = 10^4$. The rates and number of cargos are such that $w = 0.4$, $p = 0.6$, $k = 0.7$, $\alpha = k/5$, (a) $n_2 = 2$ and (b) $n_2 = 15$. *Insert* in Fig.(2a) for $T^\infty = 10^5$ at a region of low motor densities shows that the behavior of $v_m$ in this example remains essentially the same as $T^\infty$ increases 10 fold suggesting that stationary conditions have been achieved in the course of the simulations.

**Figure 3** - An enlarge view of the behavior of $v_m$ as a function of $\rho$ and the rate $p$ chosen such that $p = 1 - w$, for (a) $k = 0.7$, $\alpha = k/5$ and (b) $k = \alpha = 0.7$. Notice that although not necessary, such a relation between $p$ and $w$ offers a better view of the different possibilities for cargo´s behavior for $p \neq w$.

**Figure 4** - Cargo displacement $d(t)$ as function of time $t$, at specific values of $\rho$ chosen from the velocity profile in Fig.(2a) for $N = 100$, $T^\infty = 10^4$ and (i) $\rho = 0.05$ (ii) $\rho = 0.19$ (iii) $\rho = 0.50$ (iv) $\rho = 0.91$. The *inserts* in (i) and (ii) illustrate the magnitude of the fluctuations of $d(t)$ within the respective regions of motor densities.

**Figure 5** - Average cluster size distribution at the corresponding points (i), (ii), (iii) and (iv) of Fig.(2a), for $N = 100$ and $T^\infty = 10^4$.

**Figure 6 -** Cluster dynamics and a microscopic view of a shock wave. Cluster and cargo present a drift toward the plus end whereas motors move in the opposite direction.

**Figure 7 -** The behavior of the gap $g_N(\varepsilon, \rho)$ as a function of $\rho$, for $\varepsilon = 0.01$ and $N$ as indicated.

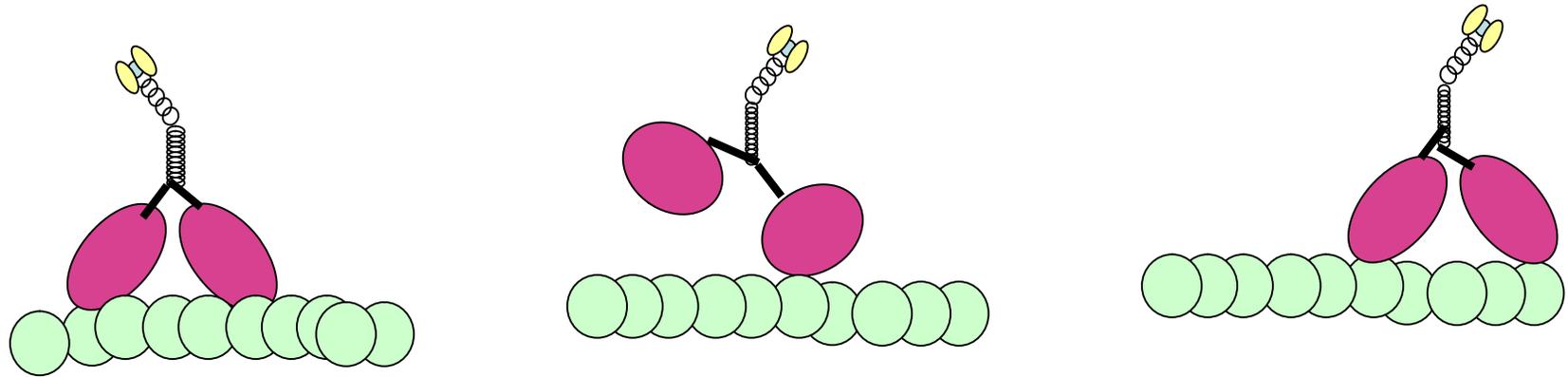

Figure 1(a)

10 →<sup>k</sup> 01

## Figure 1(b)

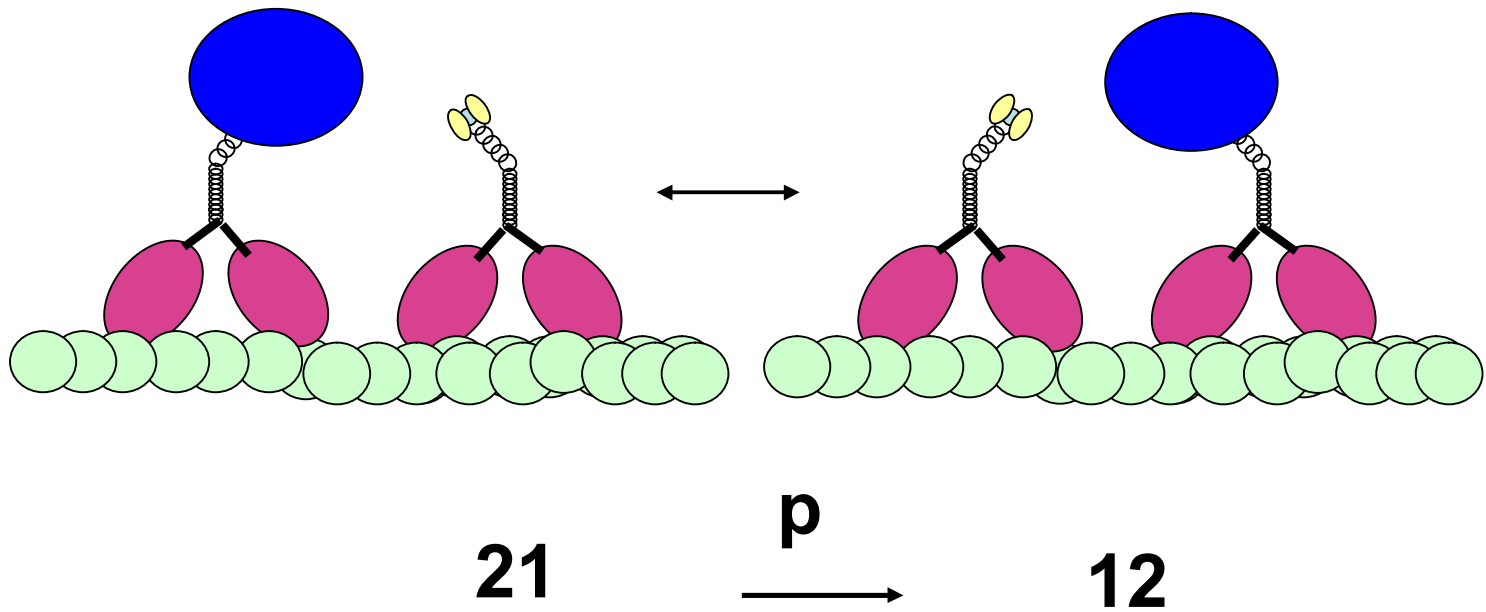

hopping

21 $\xrightarrow{p}$ 12

Figure 1(c)

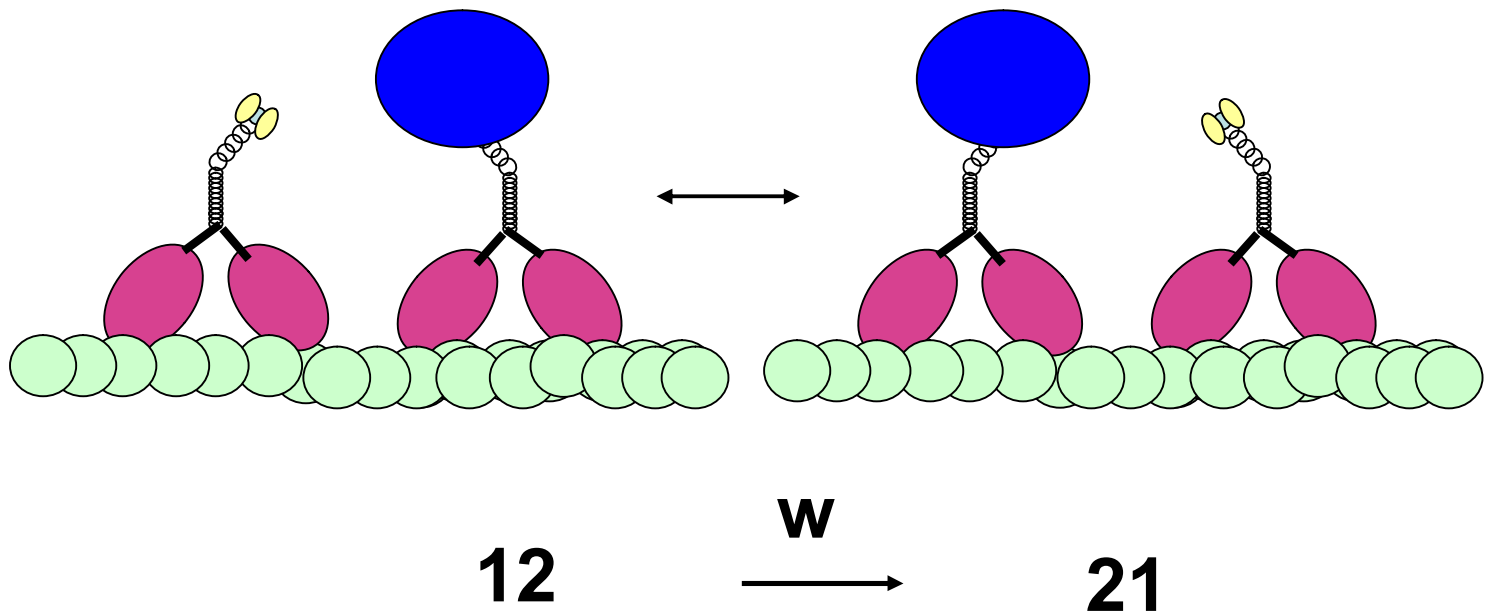

hopping

12 →w 21

## Figure 1(d)

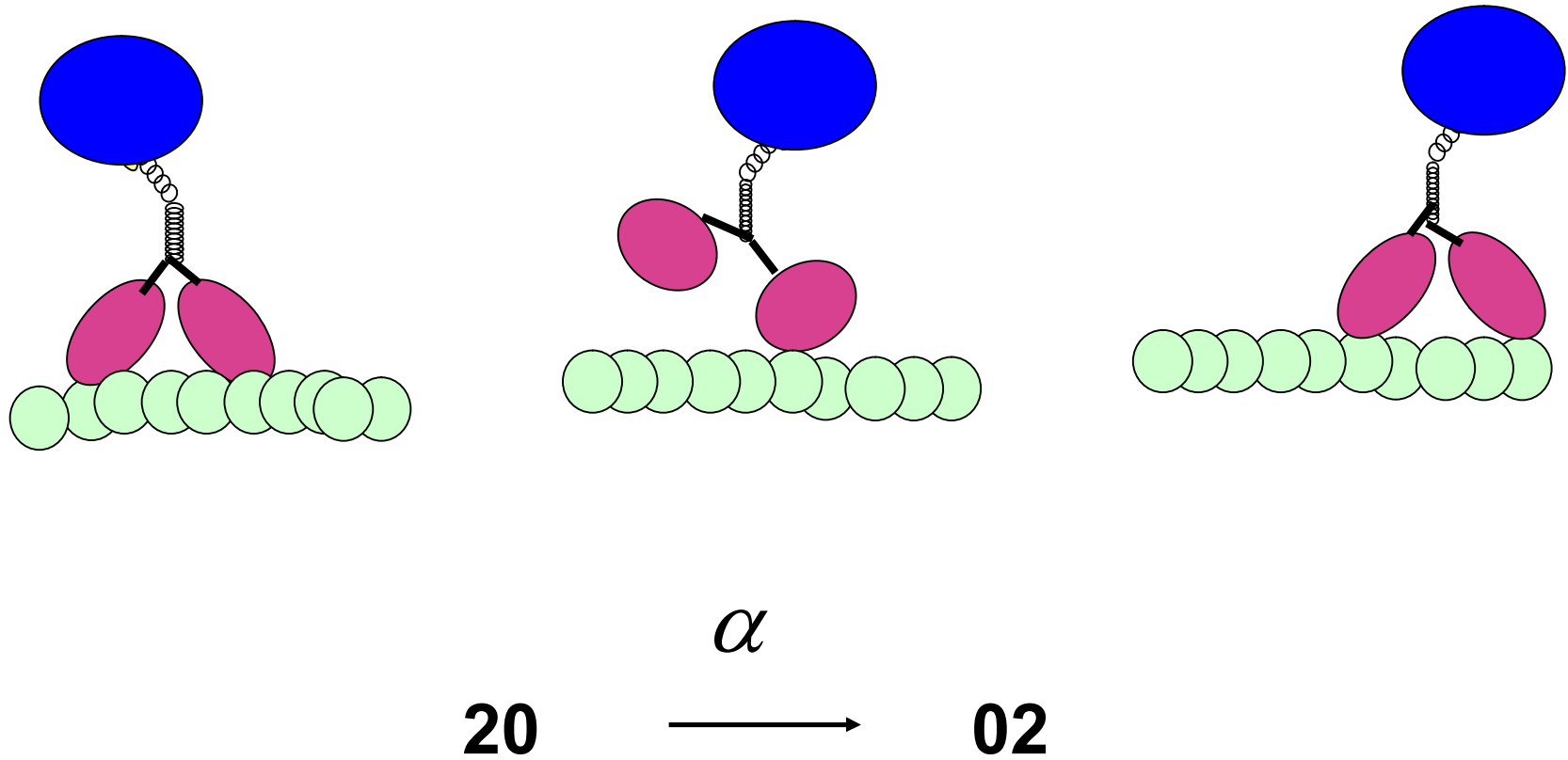

$$20 \xrightarrow{\alpha} 02$$

Fig. 2

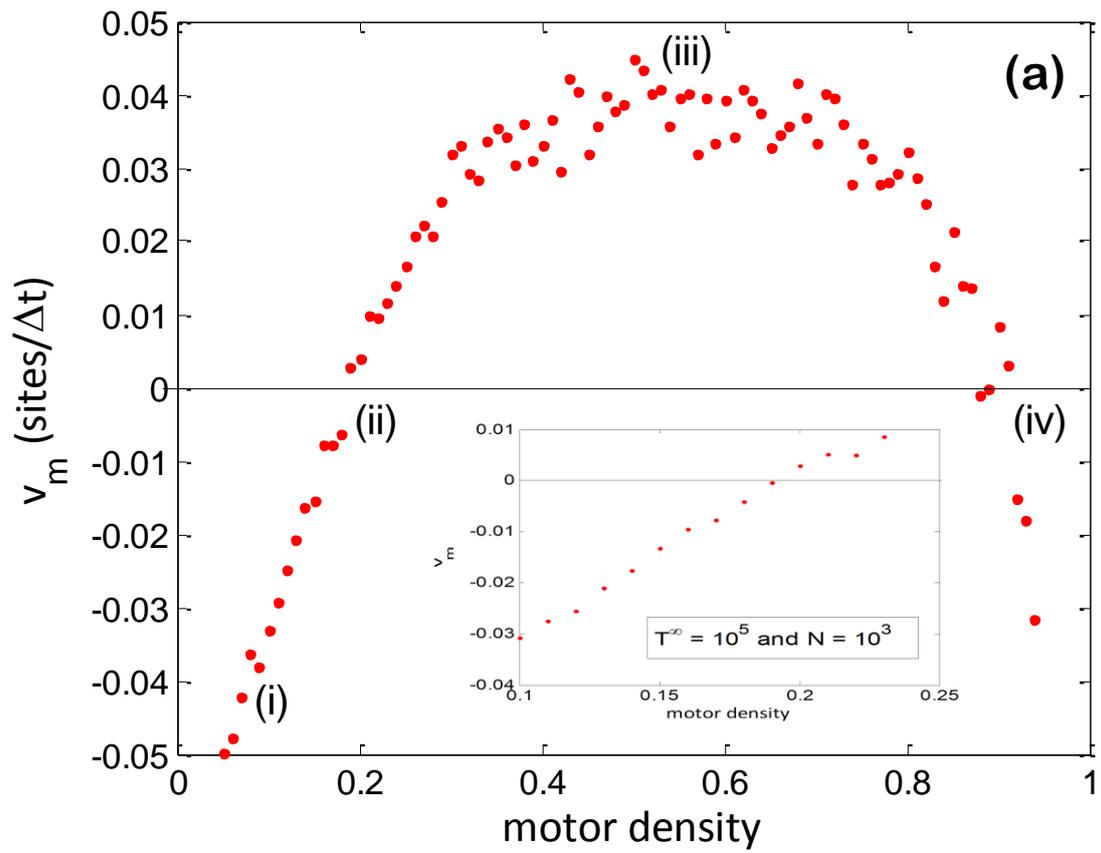

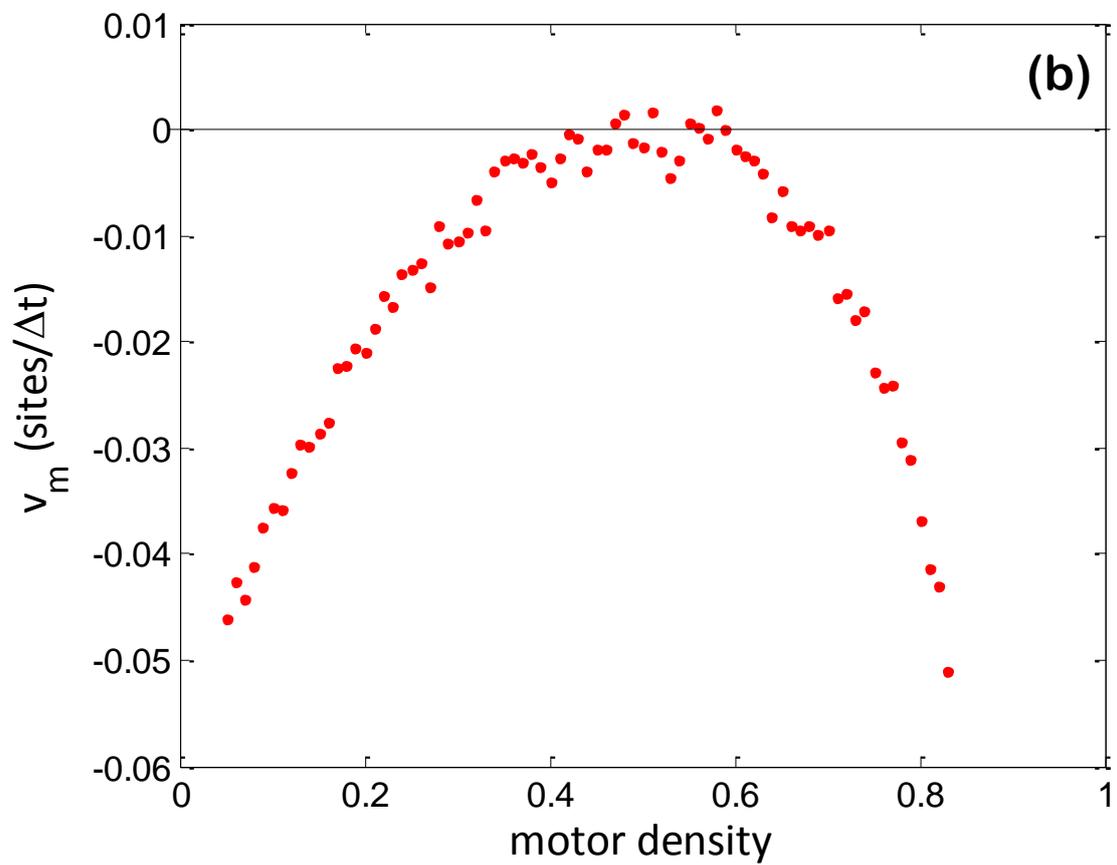

Fig. 3

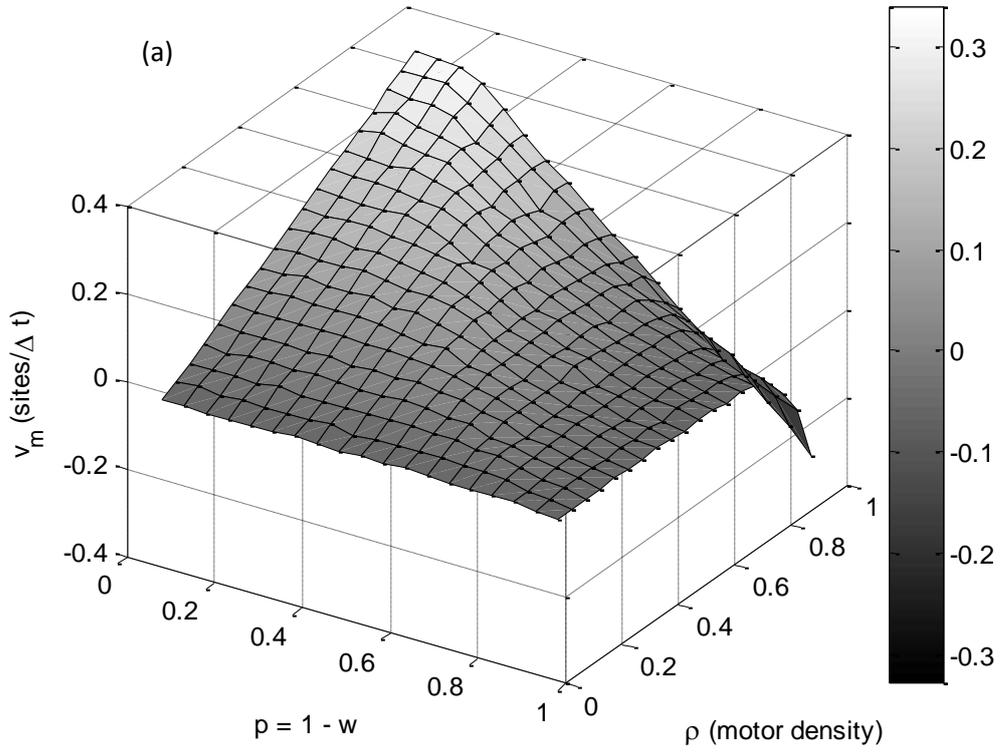

(a)

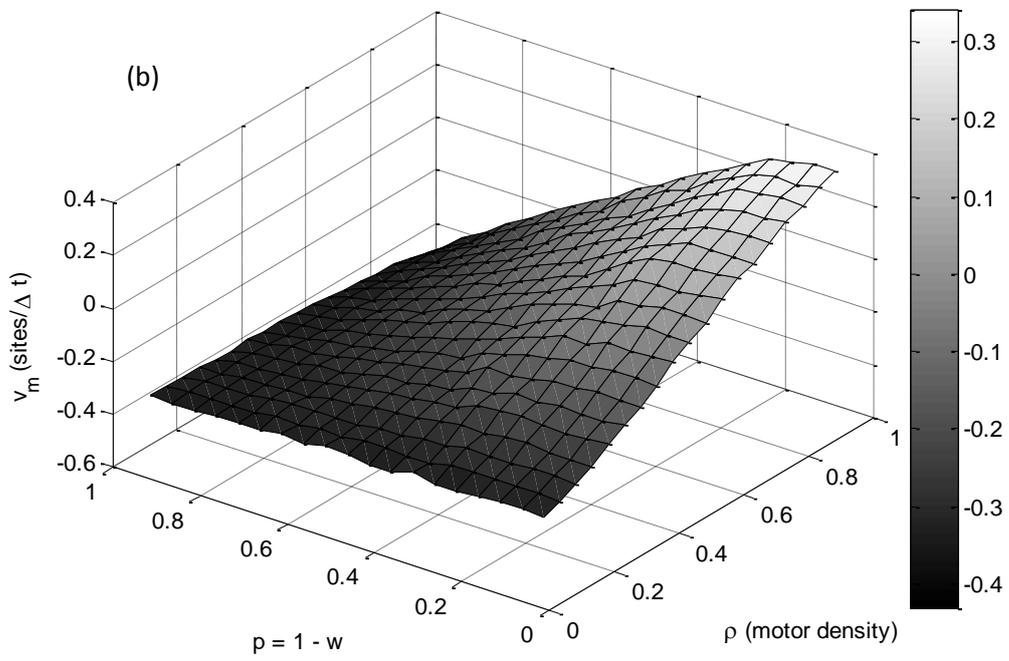

(b)

Fig. 4

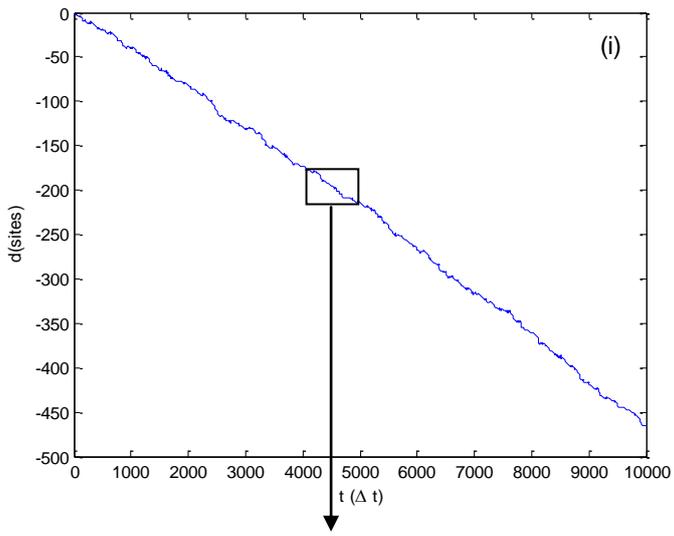
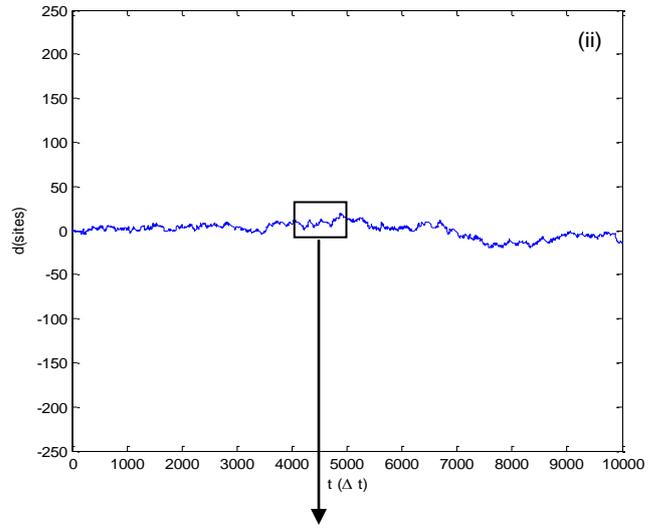
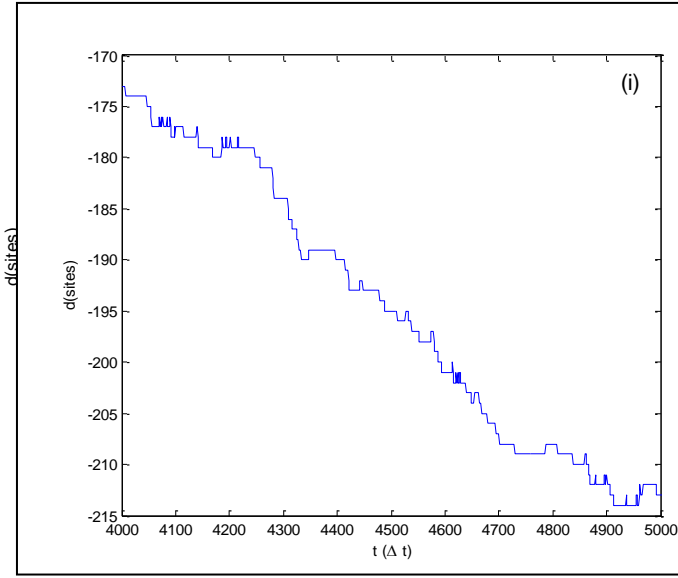
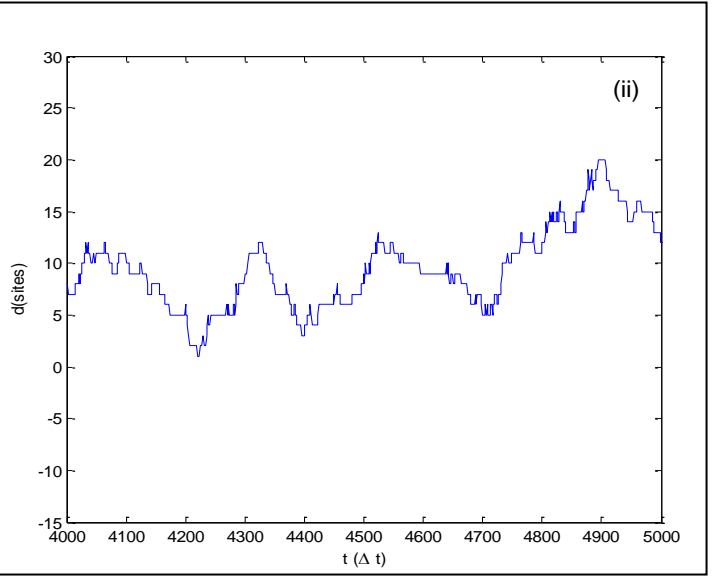
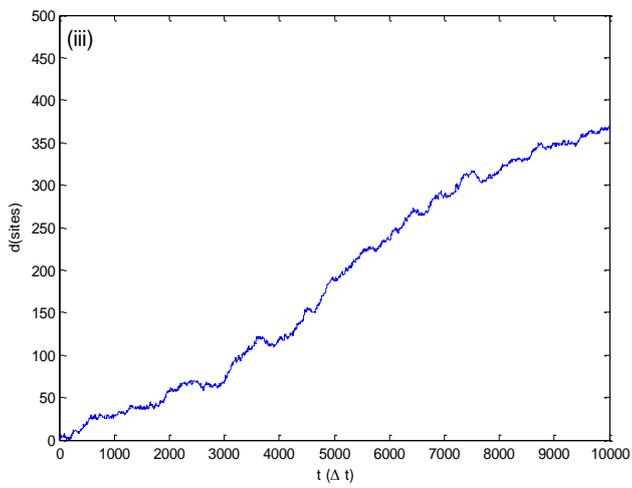
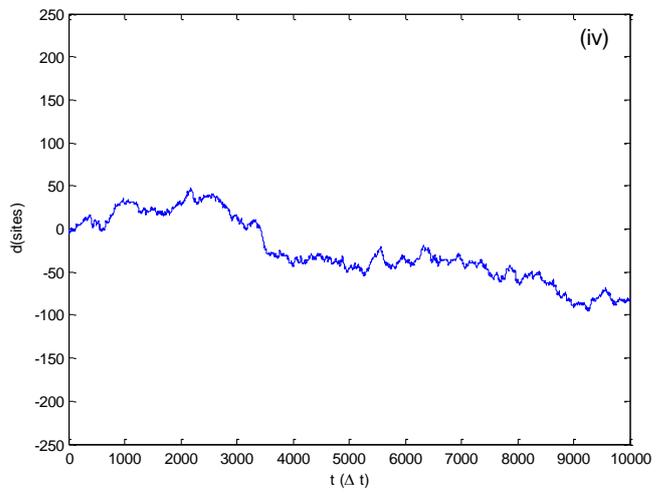

Fig. 5

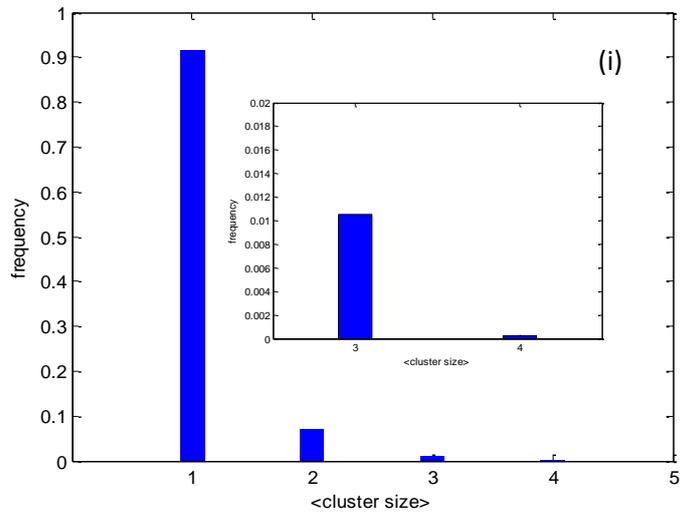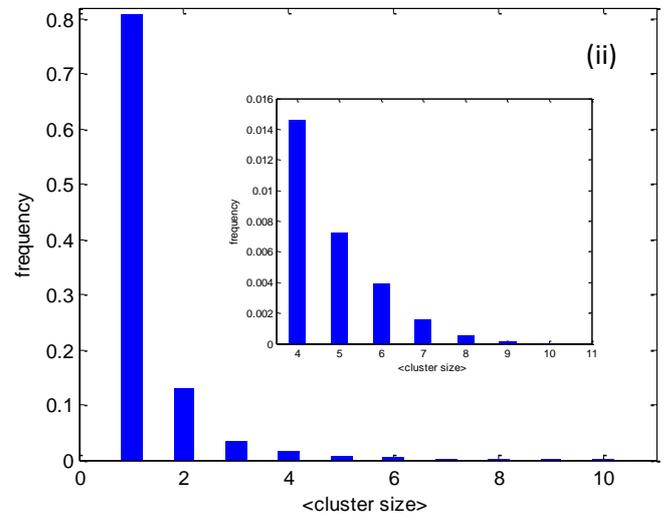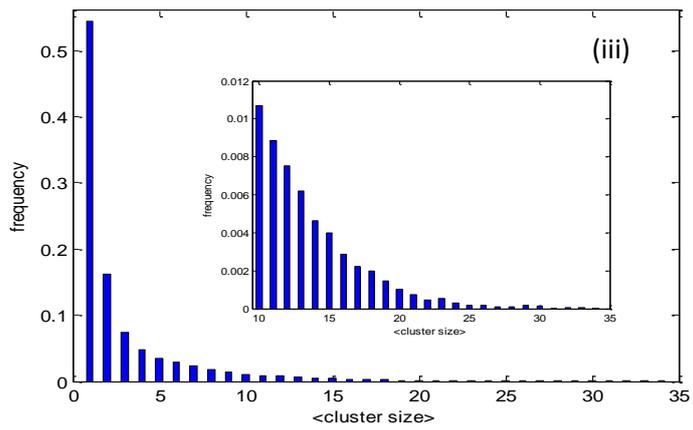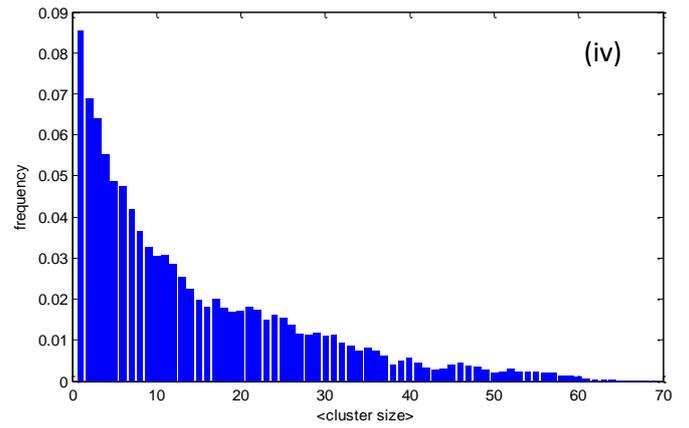

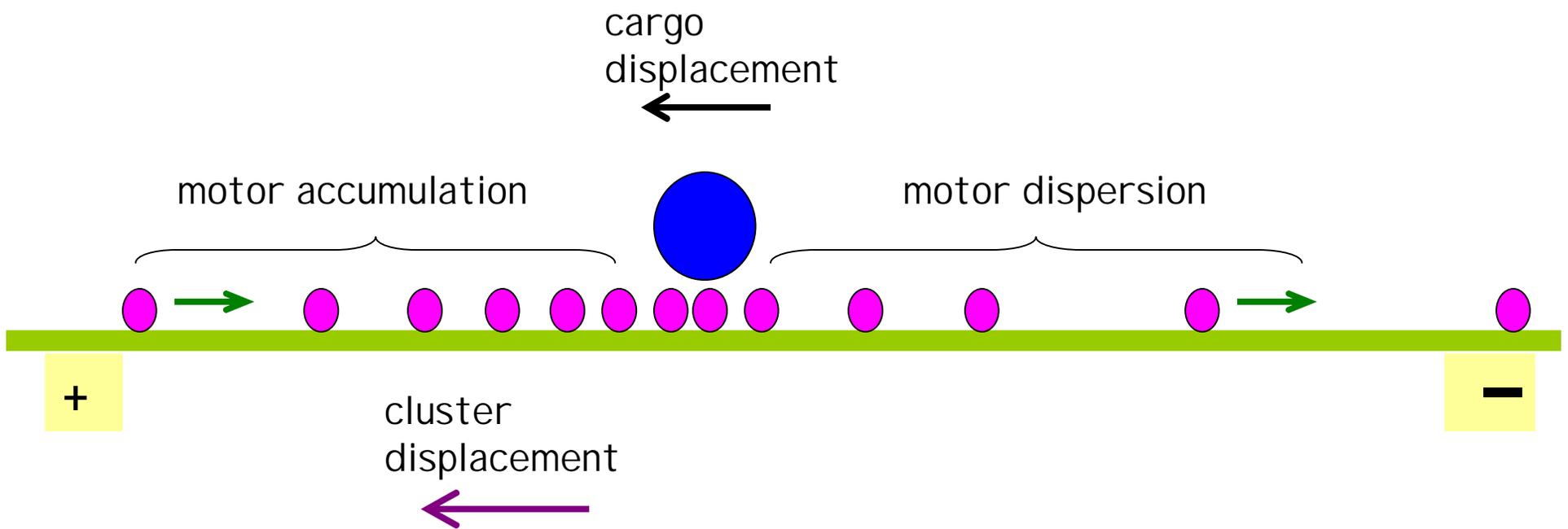

Fig. 6

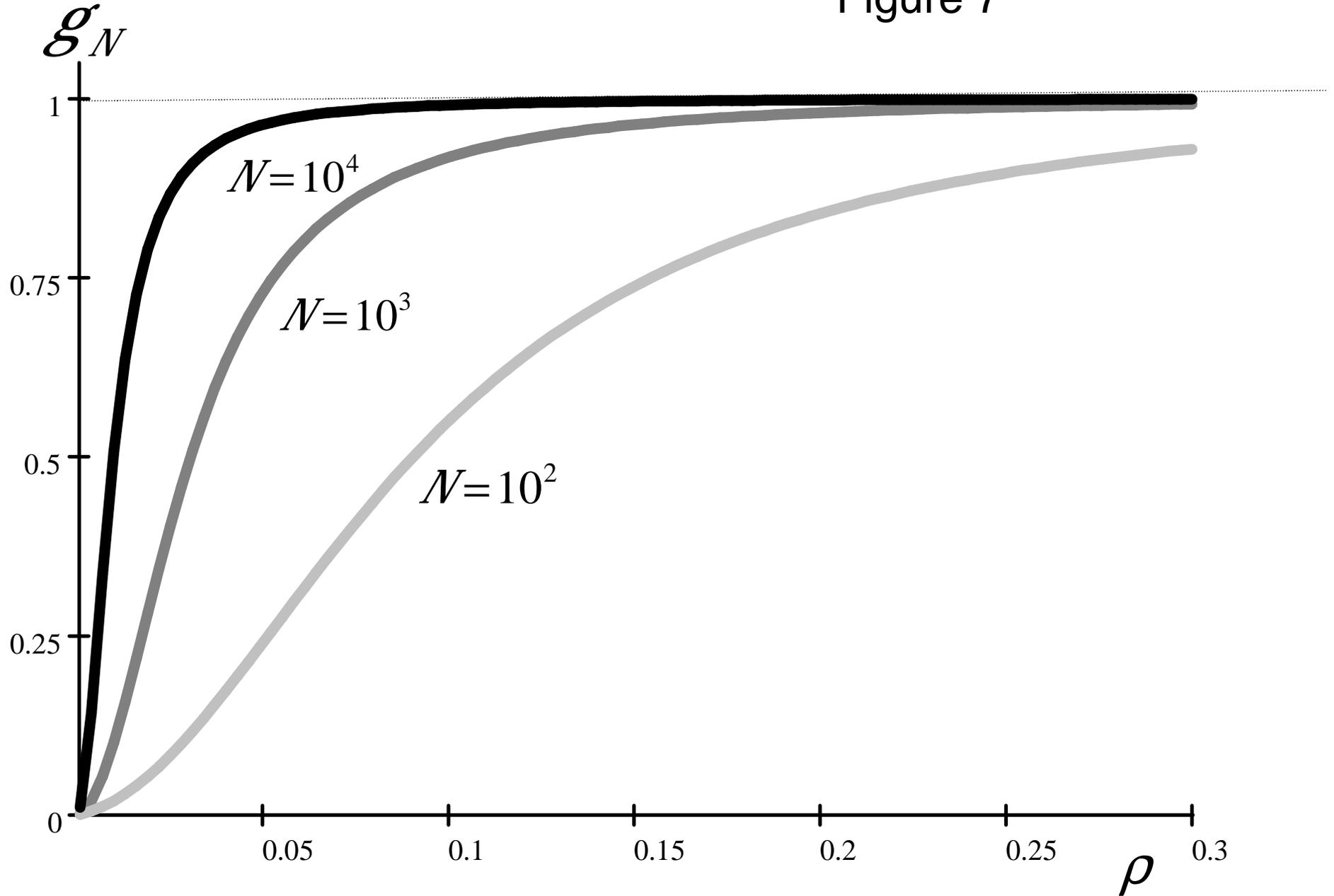

Figure 7